\documentclass[useAMS,usenatbib]{mn2e}

\usepackage{epsfig}
\usepackage{amsmath}
\usepackage{amssymb}
\usepackage{natbib}
\usepackage{threeparttable} 

\usepackage{rotating}

\usepackage[hyphens]{url}
\usepackage{hyperref}

\usepackage{epstopdf}
\usepackage[dvips]{color}

\usepackage{booktabs,array}

\usepackage{color}

\newcommand{\swiftj}{Swift J1357.2--0933}

\newcommand{\MAXI}{MAXI J1659-152}

\def\lesssim{\mathrel{\hbox{\rlap{\hbox{\lower4pt\hbox{$\sim$}}}\hbox{$<$}}}}
\let\la=\lesssim
\def\gtrsim{\mathrel{\hbox{\rlap{\hbox{\lower4pt\hbox{$\sim$}}}\hbox{$>$}}}}
\let\ga=\gtrsim

\def\lx{L_\mathrm{X}}

\def\Nh{$N_{\rm H}$}

\def\chis{$\chi^{2}$}

\newcommand{\swift}{\textit{Swift}}
\newcommand{\xmm}{\textit{XMM-Newton}}

\newcommand{\lum}{\mathrm{erg~s}^{-1}}
\newcommand{\flux}{\mathrm{erg~cm}^{-2}~\mathrm{s}^{-1}}

\newcommand{\nh}{\mathrm{cm}^{-2}}

\title[\swiftj\ in quiescence]{\swiftj: the faintest black hole?}
\author[M. Armas Padilla et al.]
{M. Armas Padilla$^{1,2}$\thanks{e-mail: montserrat.a.p.w@gmail.com}, 
R. Wijnands$^{1}$,
N. Degenaar$^{3}$,
T. Mu\~noz-Darias$^{2}$,
J. Casares$^{4,5}$
\newauthor and R. P. Fender$^{2}$
\\
$^{1}$Anton Pannekoek Institute, 
University of Amsterdam, 
Postbus 94249, 1090 GE Amsterdam, The Netherlands\\
$^{2}$University of Oxford, Department of Physics, Astrophysics, Keble Road, Oxford, OX1 3RH, United Kingdom\\
$^{3}$Department of Astronomy, University of Michigan, 500 Church Street, Ann Arbor, MI 48109, USA\\
$^{4}$Instituto de Astrof\'isica de Canarias (IAC), V\'ia L\'actea s/n, La Laguna 38205, S/C de Tenerife, Spain\\
$^{5}$Departamento de Astrof\'isica, Universidad de La Laguna, La Laguna, E-38205, S/C de Tenerife, Spain
}

\begin{document}

\date{DRAFT VERSION}

\pagerange{\pageref{firstpage}--\pageref{lastpage}} \pubyear{0000}

\maketitle

\label{firstpage}

\begin{abstract} 
\swiftj\ is a confirmed very faint black hole X-ray transient and has a short estimated orbital period of $2.8$~hr. We observed \swiftj\ for $\sim$50~ks with \xmm\ in 2013 July during its quiescent state. The source is clearly detected at a 0.5--10~keV unabsorbed flux of $\sim3\times10^{-15}~\flux$. If the source is located at a distance of 1.5~kpc (as suggested in the literature), this would imply a luminosity of $\sim8\times10^{29}~\lum$, making it the faintest detected quiescent black hole LMXB. This would also imply that there is no indication of a reversal in the quiescence X-ray luminosity versus orbital period diagram down to 2.8~hr, as has been predicted theoretically and recently supported by the detection of the 2.4~hr orbital period black hole \MAXI\ at a 0.5--10~keV X-ray luminosity of $\sim1.2\times10^{31}~\lum$. However, there is considerable uncertainty in the distance of \swiftj\ and it may be as distant as 6.3~kpc. In this case, its quiescent luminosity would be $\lx\sim1.3\times10^{31}~\lum$, i.e., similar to \MAXI\ and hence it would support the existence of such a bifurcation period. We also detected the source in optical at $r^\prime\sim$22.3 mag with the Liverpool telescope, simultaneously to our X-ray observation. The X-ray/optical luminosity ratio of \swiftj\ agrees with the expected value for a black hole at this range of quiescent X-ray luminosities.

\end{abstract}


%%%%%%%%%%%%
% INTRODUCTION
%%%%%%%%%%%%

\section{Introduction}\label{sec:intro}

Transient low-mass X-ray binaries (LMXBs) occasionally experience outburst events, caused by a sudden increase in the mass-accretion onto the compact object, a black hole (BH) or a neutron star (NS), with a resultant large increase in the X-ray luminosity. However, most of the time these sources remain in a dim, quiescent state where little or no accretion takes place. Quiescent BH binaries are systematically fainter than NS binaries, especially when comparing systems with similar orbital periods (\citealt{Narayan1997, Kong2002}). While most NS systems in quiescence have 0.5--10~keV luminosities $\lx>$~$10^{31}$ erg s$^{-1}$, BH X-ray transients can be as faint as $\lx\sim10^{30-31}$ erg s$^{-1}$. 

The low quiescent luminosities of BH LMXBs are explained invoking radiation from an advection-dominated accretion flow \citep[ADAF;][]{Narayan1997,Narayan2008}, where most of the accretion energy is eventually advected through the event horizon rather than being radiated away. In NS systems, on the other hand, all the advected energy is expected to be radiated at the neutron star surface, resulting in a much higher radiative efficiency even if the inner accretion flow is an ADAF. An alternative scenario might be the jet-dominated states in quiescence, in which the majority of the     accretion power is release in the form of radio jets rather than dissipated in the accretion flow through X-ray emission \citep{Fender2003}.  The difference in quiescent X-ray luminosity between NSs and BHs is explained by the considerable reduced jet contribution in quiescent NSs with respect to the 'radio-loud' BHs \citep{Fender2003,Gallo2006,Gallo2007,Migliari2006,Kording2007}

\citet{Menou1999} predicted a positive correlation between the orbital period and the quiescent X-ray emission based in the ADAF model. This correlation is expected to break (and reverse) at very short periods, when mass transfer switches from being driven by the expansion of the donor star to shrinkage of the orbit due to gravitational wave emission. So far, this regime of very short orbital periods has been very poorly explored \citep[e.g;][]{Homan2013}, since there are only a few BH transients with short orbital periods known.

\swiftj\ is a BH LMXB in a estimated $2.8\pm0.3$~hours orbit \citep[][]{Corral-Santana2013}. It was discovered in outburst on 2011 January 28 with Swift's Burst Alert Telescope (BAT; \citealt{Barthelmy2005,Krimm2011}). The outburst lasted 7 months and reached a 2--10~keV X-ray peak luminosity of $\lx^{\rm peak}\sim10^{35}~\lum$, for an assumed distance of 1.5 kpc  \citep{ArmasPadilla2013a}. This makes \swiftj\ a confirmed BH in a very faint X-ray binary; these are X-ray binaries that have $\lx^{\rm peak}<10^{36}~\lum$ (\citealt{Wijnands2006}). 

Based on photometry of Archival Sloan Digital Sky Survey (SDSS) data (obtained prior to the outburst), it was suggested that \swiftj\ has an M4 companion star and is located at a distance of
$D\sim$1.5~kpc \citep[][]{Rau2011,Corral-Santana2013}. However, \citet[][]{Shahbaz2013} found that the quiescent optical light is not dominated by the companion star and instead set a distance in the range $D$=0.5-6.3~kpc.

We present the results of a \xmm\ observation of \swiftj\ during quiescence and simultaneous optical data obtained with the Liverpool observatory.

\section{Observation, data analysis and results}\label{sec:data}
The \xmm\ satellite \citep{Jansen2001} pointed to \swiftj\ for 50~ks on 2013 July 10 (Observation ID 0724320101).
The European Photon Imaging Camera (EPIC) consist of two MOS cameras \citep{Turner2001} and one PN camera \citep{Struder2001}, which were operated in full-frame window imaging mode with the thin optical blocking filter applied. We used the Science Analysis Software (SAS, v.13.0) to carry out the data reduction and obtain the scientific products.
%%%%%%%%%%
%%FIGURE
%%%%%%%%%%

\begin{figure}
\begin{center}
\includegraphics[angle=0,width=\columnwidth]{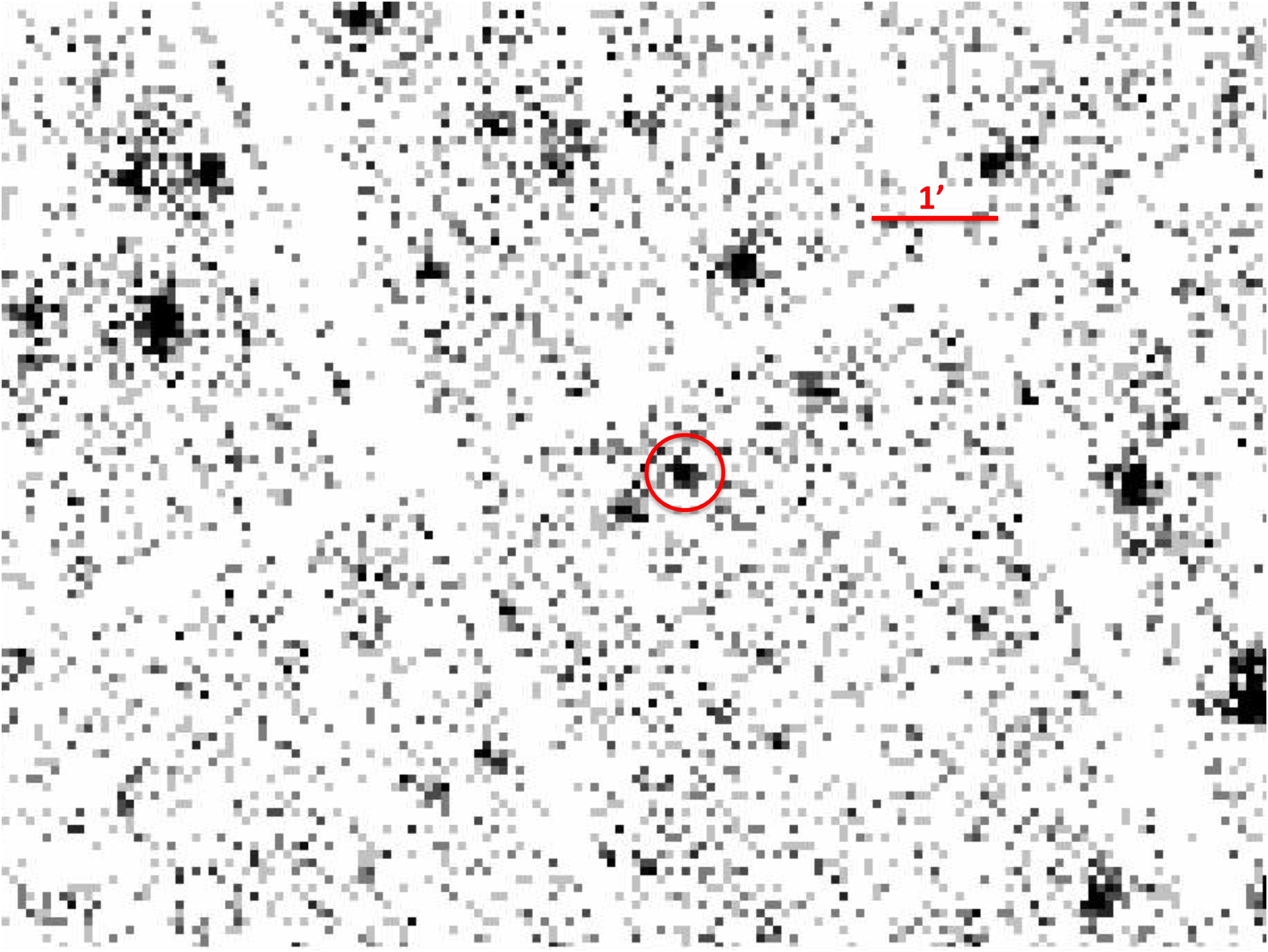}\\
\caption{\xmm\ X-ray image (0.3--12~keV) of the field around \swiftj\. To create this image we combined the data obtained with all three detectors (both MOS cameras and the PN camera). \swiftj\ is clearly detected as indicated by the red circle.}
\label{fig:ds9}
\end{center}
\end{figure}

%%%%%%%%%%%%%%%%%%%%%
In order to check for possible episodes of background flaring we examined the high-energy count rates ($>$10 keV and 10--12 keV for the MOS and PN cameras, respectively). The count rates were consistently very low throughout the observation, and therefore, not affected by any episode of background flaring. Hence, we used all data during our analysis. \swiftj\ is clearly detected in the combined 0.3-12~keV image of the three cameras, which is shown in Fig. \ref{fig:ds9}. We extracted events using a circular region with a radius of $15''$ centred on the source position (RA = 13:57:16.8, Dec. = --09:32:39, J2000, $0.3''$ error radius; \citealt{Rau2011}), and a circular region with a radius of $45''$ covering a source-free part of the CCDs to extract background events. The 0.3-12~keV average net source count rate is (1.39 $\pm$ 0.27)$\times 10^{-3}$~counts s$^{-1}$ for the PN. Very limited number of photons were recorded with the MOS cameras ($<5 \times 10^{-4}$~counts s$^{-1}$), which did not allow us to perform a spectral analysis. Therefore we only report on the spectral analysis of the PN data. The number of detected photons was insufficient to perform any variability study. 

\begin{figure}
\begin{center}
\includegraphics[angle=-90,width=\columnwidth]{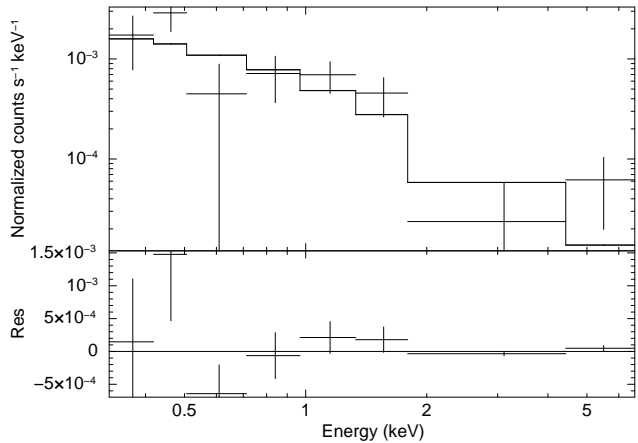}\\
\caption{The 0.3--10 keV PN X-ray spectrum (top) and residuals (bottom). The solid line represents the best fit with a simple power-law model. }
\label{fig:spec}
\end{center}
\end{figure}

We generated the spectrum and the light curve of the PN data, as well as the response files, following the standard analysis threads\footnote{http://xmm.esac.esa.int/sas/current/documentation/threads/}. The spectral data were grouped to contain a minimum of 15 photons per bin and fit in the 0.3--10~keV energy range using XSpec \citep[v.12.8;][]{xspec}. In \citet{ArmasPadilla2014a} it was shown that the \swiftj\ spectra is very little affected by absorption and it was not possible to constrain the interstellar absorption parameter (values as low as  $\sim 10^{12}\nh$ are found). These \Nh\ values are so low that they do not have a measurable influence on our spectral modelling and therefore absorption is not included in our fits. Indeed, we note that fixing the \Nh\ to $1.2 \times 10^{20}\nh$, as suggested by \citet{Krimm2011a}, has no influence in our best fit parameters. The spectrum (see Fig.\ref{fig:spec}) can be modelled with a single component model; a black body, a multi-color disk black body  and a power-law model provided equally good fits, with a p-value \textgreater 0.1 in all cases (see Table \ref{tab:results} for spectral results). In this paper we report on the results obtained using the power-law model, since it is the standard model used to fit BH spectra in quiescence \citep[e.g;][]{McClintock2006, Corbel2006, Plotkin2013}.

The obtained photon index is $\Gamma=2.1\pm0.4$ with a reduced \chis of 1.37 for 6 degrees of freedom (p-value 0.22). The unabsorbed 0.5-10~keV flux is $(3.2^{+1.1}_{-0.7})\times10^{-15}~\lum$ and $(1.6^{+1.1}_{-0.7})\times10^{-15}~\lum$ in the 2--10~keV band. The errors on the spectral parameters reflect 1$\sigma$ uncertainties and the flux errors have been calculated following the procedure presented by \citet{Wijnands2004}.

%%%%%%%%%%%%%%%%%%%%%%%%%%%
%   TABLE
%%%%%%%%%%%%%%%%%%%%%%%%%%%

\begin{table}
\begin{center}
\caption{Results from fitting the spectral data.}
\label{tab:results}
\begin{threeparttable}
\begin{tabular}{l c c}
\hline
Model & $\Gamma$, $kT$[keV] & \chis (dof)  \\
\hline
\textsc{powerlaw} & $2.1 \pm 0.4 $ & 1.37 (6) \\
\textsc{bbodyrad} & $0.31 \pm 0.07 $ & 1.58 (6) \\
\textsc{diskbb} & $0.5 \pm 0.2 $ & 1.39 (6) \\
\hline
\end{tabular}
%\begin{tablenotes}
%\item[Note]{To calculate the parameter errors, we fixed the instrumental features components. }
%\end{tablenotes}
\end{threeparttable}
\end{center}
\end{table}

\subsection{Optical observations}\label{subsec:opt}

We observed \swiftj\ with the IO:O camera attached to the 2m Liverpool Telescope in La Palma, Spain \citep{Steele2004}. Our observations were carried out $\sim$~30 months after the outburst on 2013 Jul 9 ($4\times300$s), Jul 10 ($4\times300$s; these were simultaneous with our \xmm\ observation) and Jul 12 ($3\times300$s), using the Sloan $r^\prime$ filter. The data were bias and flat-field corrected and fluxes were subsequently extracted using standard \textsc{iraf} routines for aperture photometry. Following \citet{Shahbaz2013} we calibrated the field using the Sloan Digital Sky Survey (SDSS) star J135716.43-093140.1 located only $\sim 1'$ away from the object. By combining the exposures obtained each night we measure averaged magnitudes of $r^\prime$=  $22.18 \pm 0.09$, $22.29 \pm 0.08$ and $22.24 \pm 0.10$, on Jul 9, Jul 10 and Jul 12, respectively. Given the very low extinction in the line-of-sight this values were not corrected from reddening.

The source was detected prior to its 2011 outburst by the SDSS at $r^\prime$=  $21.99 \pm 0.14$ \citep{Rau2011}. \citet{Shahbaz2013} reported $r^\prime= 21.54 \pm 0.35$ on 2012 Apr, 24 months after the outburst. In this study the large error reflects the variability (rms) of the light-curve, since strong flaring behaviour on time scales of seconds to minutes was present. Our eleven 300s exposures also show significant scatter with magnitudes ranging from $23.00 \pm 0.29$ to $21.78 \pm 0.1$ with an average value of $22.16 \pm 0.52$ ($22.29 \pm 0.31$ for Jul 10), where the latter error reflects the rms variability. Given the very similar night averaged magnitudes, this flaring behaviour seems to be washed out on time scales of $\sim 20$ min. We conclude that during our \xmm\ observation the source is fully consistent with the pre-outburst magnitude and, on average, slightly fainter than during the \citet{Shahbaz2013} observations.

%%%%%%%%%%
%%FIGURE
%%%%%%%%%%

%%%%%%%%%%%%%%%%%%%%%

%%%%%%%%%%
%%FIGURE
%%%%%%%%%%

\begin{figure}
\begin{center}
\includegraphics[angle=0,width=\columnwidth]{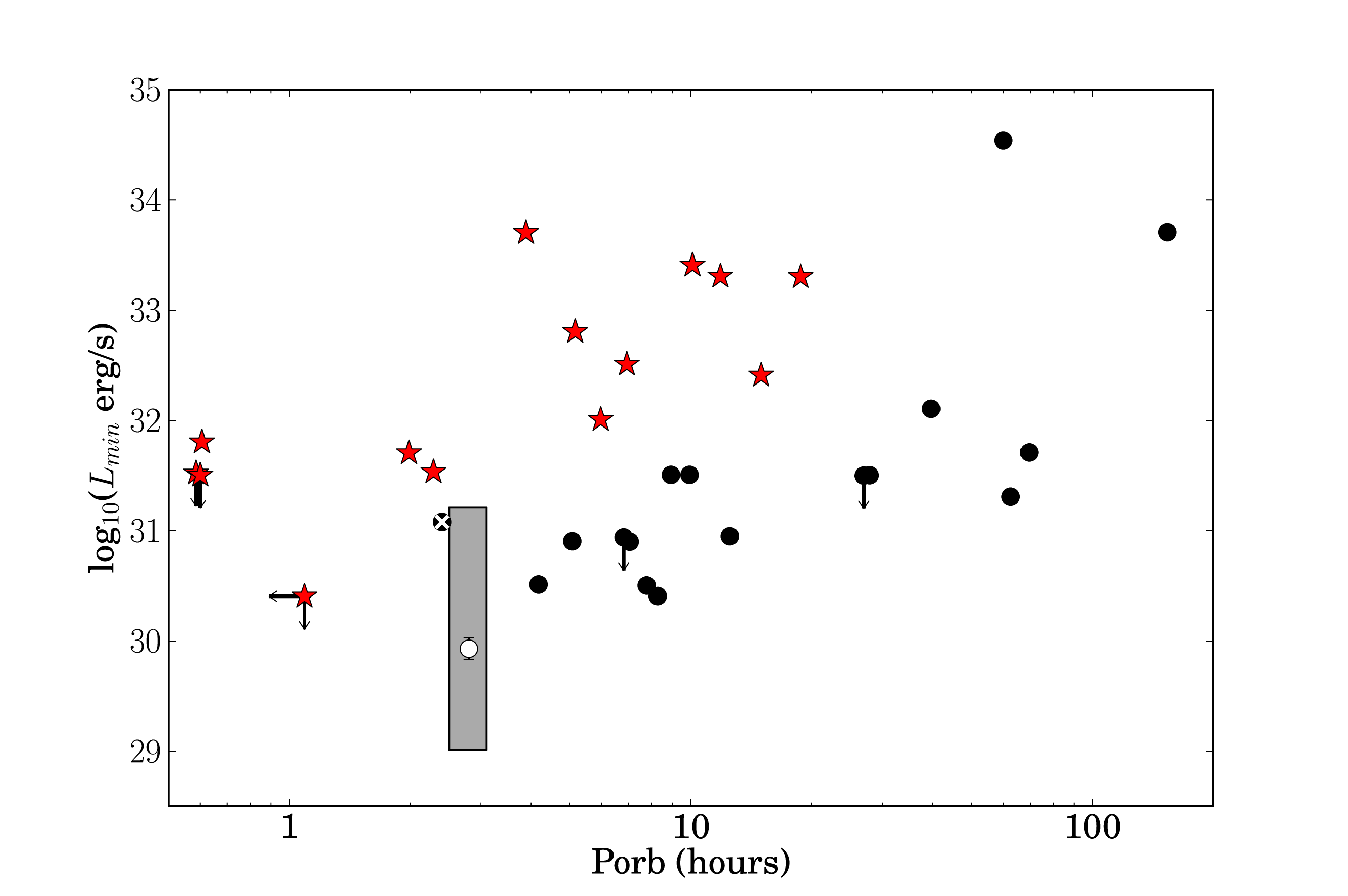}\\
\caption{ Quiescent 0.5-10~keV luminosities versus the orbital period ($P_{\rm orb}$) for NS (red stars) and BH (circles) X-ray binaries. Arrows represent upper limits on luminosity or $P_{\rm orb}$. The luminosity of \swiftj\ assuming a distance of 1.5~kpc is plotted with a white circle. The grey area reflects the uncertainty in the distance to the source and $P_{\rm orb}$. The luminosity of \MAXI\ is plotted with a crossed black circle. Based on data presented in \citet{Gallo2008}, \citet{Rea2011}, \citet{Reynolds2011}, \citet{Homan2013},    \citet{Yang2013} and \citet{Casares2013}. (A color version of this figure is available in the online journal.)}
\label{fig:Porb_L}
\end{center}
\end{figure}

%%%%%%%%%%%%%%%%%%%%%

\section{Discussion}\label{sec:disc}

We present the \xmm\ detection of the very faint BH LMXB \swiftj\ in its quiescent state. The X-ray spectrum can be modelled by an power law with a photon index of $2.1\pm0.4$, similar to what is reported for other BH transients in quiescence \citep[e.g;][]{Kong2002,Corbel2004,Wijnands2012,Homan2013,Plotkin2013}. The 0.5--10~keV unabsorbed flux is $(3.2^{+1.1}_{-0.7})\times10^{-15}~\flux$, which is $\sim$ 2 orders of magnitude lower than the \swift/XRT upper-limit reported by \citep{ArmasPadilla2013a}. 

During the last \swift/XRT detection of \swiftj\ reported by \citep{ArmasPadilla2013a}, the source spectrum was also described by a power-law with a photon index of $2.1^{+1.8}_{-0.9}$. Despite the large errors on the photon indices, the source spectra are thus consistent with being the same in quiescence and at the end of the outburst when the source was about 100 times brighter. This behavior is very similar to that found by \citet{Plotkin2013} for a sample of BH transients. They reported that below a luminosity of $10^{33-34}~\lum$, the sources did not evolve in spectral shape anymore with the photon index plateauing to an average 2.08. \swiftj\ thus fits within this picture.

Assuming a distance of 1.5~kpc \citep{Rau2011, Corral-Santana2013} the corresponding X-ray luminosity would be $8.5^{+5.5}_{-2.6}\times10^{29}~\lum$, which would make \swiftj\ the faintest BH yet in quiescence (below the $\sim3\times10^{30}\lum$ luminosity of GS 2000+250, XTE J1650-500 and A0620-00; \citealt{Gallo2008, Garcia2001}, see Fig.~\ref{fig:Porb_L}). However, this distance estimate is very uncertain and \citet{Shahbaz2013} actually suggested that it could range between 0.5 and 6.3 kpc. If the source were at $D=0.5$ kpc, it would even be fainter than what we have stated above. On the other hand, if we assume a distance of 3 kpc, the luminosity would be  $\sim3\times10^{30}\lum$, comparable to the faintest BH transients detected in quiescence. Finally, if we adopt the largest proposed distance of 6.3~kpc, the luminosity would be $\sim1.5\times10^{31}~\lum$. This value is similar to that of the 2.4 h orbital period system \MAXI\ with $\sim1.2\times10^{31}(d/6~\rm kpc)^{2}~\lum$, where the distance (\textit{d}) ranges between 4.5 and 8.5~kpc \citep[see crossed black circle in Fig. \ref{fig:Porb_L}; ][]{Kuulkers2013, Homan2013}. 

The uncertainty in the distance towards \swiftj\ is indicated by the grey area in Fig.~\ref{fig:Porb_L}. The large uncertainty makes it difficult to draw any conclusion on the behaviour of X-ray luminosity with orbital period. \citet{Menou1999} predicted that the quiescent luminosity of BH transients would decrease when the orbital period decreases, but starts to increase again for orbital periods under 5--10 hours (the bifurcation period). \citet{Homan2013} found that the BH transient \MAXI\ indeed had a rather high quiescent luminosity compared to systems with longer orbital periods, in line with Menou et al's predictions. However, if \swiftj\ is not that far away ($<$3 kpc), it would be (at most) as bright as the faintest detected quiescent BH transients, with orbital periods up to 10 hrs. This would then suggest that the bifurcation period might not exist or is located at a lower value. Also the data point of \MAXI\ would be anomalous, perhaps indicating enhanced emission due to enhanced accretion activity above the true quiescence level (Fig.~\ref{fig:Porb_L}; see also \citealt{Homan2013}). 

On the other hand, if \swiftj\ were at $D=3-6.3$~kpc, it would have a similar quiescent luminosity as \MAXI\ and then it would lend support to the existence of a bifurcation period in the range of a few hours. We stress that we cannot make strong conclusions without a more accurate distance estimate for \swiftj. In addition, \citet{Menou1999} stated that the exact behavior of the quiescent luminosity with orbital period also depends on the mass of the companion star. Since this is not known for \swiftj\ \citep{Shahbaz2013}, it adds to the uncertainties. Finally, we note that alternative scenarios (e.g. jet-dominated; see section \ref{sec:intro}) different from the ADAF might be also able to explain the observed quiescent luminosities, although it is unclear if in those scenarios the quiescent luminosity is correlated with the orbital period as well.

Our simultaneous optical observations detected \swiftj\ at a magnitude of $r^\prime\sim$22.3. This is the faintest optical level observed in this source and it is consistent with its pre-outburst brightness. It corresponds to an optical luminosity density of $\sim 1\times10^{16-17}~\rm erg ~\rm s^{-1} \rm Hz^{-1}$, in agreement with the expected value for a BH with an X-ray luminosity in the range $\sim 1\times10^{29-31}~\lum$ \citep{Russell2006}. Nevertheless, the uncertainty in the distance does not allow us to distinguish between the optical light being dominated by jet emission (as proposed by \citealt{Shahbaz2013}) or by the accretion disc \citep[Fig.4 in ][]{Russell2006}.

With our detected X-ray luminosity, we can calculate what the expected radio luminosity would be if the source follows the standard universal X-ray - radio correlation found for BH transients \citep{Corbel2003,Gallo2003}. Using the relation reported by \citet{Gallo2012} for the upper-track correlation, we find that the source should have radio luminosities of 1, 2.5, or 6.3 $\times10^{26}~\lum$ for a distance of 1.5, 3 and 6.3 kpc, respectively. These luminosities correspond to radio flux densities of 4.3, 2.7 and 1.55 $\rm \mu$Jy at 8.46~GHz, respectively. These predicted radio fluxes are very low, making the source difficult to detect in radio during its quiescent state.

\section*{Acknowledgments}

MAP acknowledges the hospitality of the Instituto de Astrof\'isica de Canarias, where part of this work was carried out. We acknowledge the \xmm\ team for making this observation possible. RW and MAP are supported by an European Research Counsil starting grant awarded to RW. ND is supported by NASA through Hubble Postdoctoral Fellowship grant number HST-HF-51287.01-A from the Space Telescope Science Institute (STScI). TMD acknowledges funding via an EU Marie Curie Intra-European Fellowship under contract no. 2011-301355. JC acknowledges support  by the Spanish MINECO under grant AYA2010-18080, and also from the Spanish Ministry of Economy and Competitiveness (MINECO) under the 2011 Severo Ochoa Program MINECO SEV-2011-0187. This project was funded in part by European Research Council Advanced Grant 267697 4$\pi$-sky: Extreme Astrophysics with Revolutionary Radio Telescopes

%%%%%%%%%%%%%%%%%%%%%%%%%%%%%%%%%%%%%%%%%%%%%%%
\bibliographystyle{mn2e}
\bibliography{bibliography}

\begin{thebibliography}{}

\bibitem[\protect\citeauthoryear{{Armas Padilla}, Degenaar, Russell \&
  Wijnands}{{Armas Padilla} et~al.}{2013}]{ArmasPadilla2013a}
{Armas Padilla} M.,  Degenaar N.,  Russell D.~M.~M.,    Wijnands R.,  2013,
  \mnras, 428, 3083

\bibitem[\protect\citeauthoryear{{Armas Padilla}, Wijnands, Altamirano, Mendez,
  Miller \& Degenaar}{{Armas Padilla} et~al.}{2014}]{ArmasPadilla2014a}
{Armas Padilla} M.,  Wijnands R.,  Altamirano D.,  Mendez M.,  Miller J.~M.,
  Degenaar N.,  2014, Monthly Notices of the Royal Astronomical Society

\bibitem[\protect\citeauthoryear{Arnaud}{Arnaud}{1996}]{xspec}
Arnaud K.~A.,  1996, in Jacoby G.~H.,  Barnes J.,  eds, Astronomical Data
  Analysis Software and Systems V Vol.~101 of Astronomical Society of the
  Pacific Conference Series, {XSPEC: The First Ten Years}.
p.~17

\bibitem[\protect\citeauthoryear{Barthelmy, Barbier, Cummings, Fenimore,
  Gehrels, Hullinger, Krimm, Markwardt, Palmer, Parsons, Sato, Suzuki,
  Takahashi, Tashiro \& Tueller}{Barthelmy et~al.}{2005}]{Barthelmy2005}
Barthelmy S.~D.,  Barbier L.~M.,  Cummings J.~R.,  Fenimore E.~E.,  Gehrels N.,
   Hullinger D.,  Krimm H.~A.,  Markwardt C.~B.,  Palmer D.~M.,  Parsons A.,
  Sato G.,  Suzuki M.,  Takahashi T.,  Tashiro M.,    Tueller J.,  2005, \ssr,
  120, 143

\bibitem[\protect\citeauthoryear{Casares \& Jonker}{Casares \&
  Jonker}{2013}]{Casares2013}
Casares J.,  Jonker P.~G.,  2013, p.~28

\bibitem[\protect\citeauthoryear{Corbel, Fender, Tomsick, Tzioumis \&
  Tingay}{Corbel et~al.}{2004}]{Corbel2004}
Corbel S.,  Fender R.~P.,  Tomsick J.~A.,  Tzioumis A.~K.,    Tingay S.,  2004,
  The Astrophysical Journal, 617, 1272

\bibitem[\protect\citeauthoryear{Corbel, Nowak, Fender, Tzioumis \&
  Markoff}{Corbel et~al.}{2003}]{Corbel2003}
Corbel S.,  Nowak M.~A.,  Fender R.~P.,  Tzioumis A.~K.,    Markoff S.,  2003,
  Astronomy and Astrophysics, 400, 1007

\bibitem[\protect\citeauthoryear{Corbel, Tomsick \& Kaaret}{Corbel
  et~al.}{2006}]{Corbel2006}
Corbel S.,  Tomsick J.~A.,    Kaaret P.,  2006, \apj, 636, 971

\bibitem[\protect\citeauthoryear{Corral-Santana, Casares, Mu\~{n}oz Darias,
  Rodr\'{\i}guez-Gil, Shahbaz, Torres, Zurita \& Tyndall}{Corral-Santana
  et~al.}{2013}]{Corral-Santana2013}
Corral-Santana J.~M.,  Casares J.,  Mu\~{n}oz Darias T.,  Rodr\'{\i}guez-Gil
  P.,  Shahbaz T.,  Torres M.~A.~P.,  Zurita C.,    Tyndall A.~A.,  2013,
  Science, 339, 1048

\bibitem[\protect\citeauthoryear{Fender, Gallo \& Jonker}{Fender
  et~al.}{2003}]{Fender2003}
Fender R.~P.,  Gallo E.,    Jonker P.~G.,  2003, Monthly Notices of the Royal
  Astronomical Society, 343, L99

\bibitem[\protect\citeauthoryear{Gallo, Fender, Miller-Jones, Merloni, Jonker,
  Heinz, Maccarone \& {Van Der Klis}}{Gallo et~al.}{2006}]{Gallo2006}
Gallo E.,  Fender R.~P.,  Miller-Jones J. C.~A.,  Merloni A.,  Jonker P.~G.,
  Heinz S.,  Maccarone T.~J.,    {Van Der Klis} M.,  2006, Monthly Notices of
  the Royal Astronomical Society, 370, 1351

\bibitem[\protect\citeauthoryear{Gallo, Fender \& Pooley}{Gallo
  et~al.}{2003}]{Gallo2003}
Gallo E.,  Fender R.~P.,    Pooley G.~G.,  2003, Monthly Notices of the Royal
  Astronomical Society, 344, 60

\bibitem[\protect\citeauthoryear{Gallo, Homan, Jonker \& Tomsick}{Gallo
  et~al.}{2008}]{Gallo2008}
Gallo E.,  Homan J.,  Jonker P.~G.,    Tomsick J.~A.,  2008, The Astrophysical
  Journal, 683, L51

\bibitem[\protect\citeauthoryear{Gallo, Migliari, Markoff, Tomsick, Bailyn,
  Berta, Fender \& Miller‐Jones}{Gallo et~al.}{2007}]{Gallo2007}
Gallo E.,  Migliari S.,  Markoff S.,  Tomsick J.~A.,  Bailyn C.~D.,  Berta S.,
  Fender R.,    Miller‐Jones J. C.~A.,  2007, The Astrophysical Journal, 670,
  600

\bibitem[\protect\citeauthoryear{Gallo, Miller \& Fender}{Gallo
  et~al.}{2012}]{Gallo2012}
Gallo E.,  Miller B.,    Fender R.,  2012, p.~10

\bibitem[\protect\citeauthoryear{Garcia, McClintock, Narayan, Callanan, Barret
  \& Murray}{Garcia et~al.}{2001}]{Garcia2001}
Garcia M.~R.,  McClintock J.~E.,  Narayan R.,  Callanan P.,  Barret D.,
  Murray S.~S.,  2001, The Astrophysical Journal, 553, L47

\bibitem[\protect\citeauthoryear{Homan, Fridriksson, Jonker, Russell, Gallo,
  Kuulkers, Rea \& Altamirano}{Homan et~al.}{2013}]{Homan2013}
Homan J.,  Fridriksson J.~K.,  Jonker P.~G.,  Russell D.~M.,  Gallo E.,
  Kuulkers E.,  Rea N.,    Altamirano D.,  2013, The Astrophysical Journal,
  775, 9

\bibitem[\protect\citeauthoryear{Jansen, Lumb, Altieri, Clavel, Ehle, Erd,
  Gabriel, Guainazzi, Gondoin, Much, Munoz, Santos, Schartel, Texier \&
  Vacanti}{Jansen et~al.}{2001}]{Jansen2001}
Jansen F.,  Lumb D.,  Altieri B.,  Clavel J.,  Ehle M.,  Erd C.,  Gabriel C.,
  Guainazzi M.,  Gondoin P.,  Much R.,  Munoz R.,  Santos M.,  Schartel N.,
  Texier D.,    Vacanti G.,  2001, \aap, 365, L1

\bibitem[\protect\citeauthoryear{Kong, McClintock, Garcia, Murray \&
  Barret}{Kong et~al.}{2002}]{Kong2002}
Kong A. K.~H.,  McClintock J.~E.,  Garcia M.~R.,  Murray S.~S.,    Barret D.,
  2002, The Astrophysical Journal, 570, 277

\bibitem[\protect\citeauthoryear{K\"{o}rding, Migliari, Fender, Belloni, Knigge
  \& McHardy}{K\"{o}rding et~al.}{2007}]{Kording2007}
K\"{o}rding E.~G.,  Migliari S.,  Fender R.,  Belloni T.,  Knigge C.,
  McHardy I.,  2007, Monthly Notices of the Royal Astronomical Society, 380,
  301

\bibitem[\protect\citeauthoryear{Krimm, Barthelmy, Baumgartner, Cummings,
  Fenimore, Gehrels, Markwardt, Palmer, Sakamoto, Skinner, Stamatikos, Tueller
  \& Ukwatta}{Krimm et~al.}{2011}]{Krimm2011}
Krimm H.~A.,  Barthelmy S.~D.,  Baumgartner W.,  Cummings J.,  Fenimore E.,
  Gehrels N.,  Markwardt C.~B.,  Palmer D.,  Sakamoto T.,  Skinner G.,
  Stamatikos M.,  Tueller J.,    Ukwatta T.,  2011, \atel, 3138, 1

\bibitem[\protect\citeauthoryear{Krimm, Kennea \& Holland}{Krimm
  et~al.}{2011}]{Krimm2011a}
Krimm H.~A.,  Kennea J.~A.,    Holland S.~T.,  2011, \atel, 3142, 1

\bibitem[\protect\citeauthoryear{Kuulkers, Kouveliotou, Belloni, {Cadolle Bel},
  Chenevez, {D\'{\i}az Trigo}, Homan, Ibarra, Kennea, Mu\~{n}oz Darias, Ness,
  Parmar, Pollock, van~den Heuvel \& van~der Horst}{Kuulkers
  et~al.}{2013}]{Kuulkers2013}
Kuulkers E.,  Kouveliotou C.,  Belloni T.,  {Cadolle Bel} M.,  Chenevez J.,
  {D\'{\i}az Trigo} M.,  Homan J.,  Ibarra A.,  Kennea J.~A.,  Mu\~{n}oz Darias
  T.,  Ness J.-U.,  Parmar A.~N.,  Pollock A. M.~T.,  van~den Heuvel E. P.~J.,
    van~der Horst A.~J.,  2013, Astronomy \& Astrophysics, 552, A32

\bibitem[\protect\citeauthoryear{McClintock \& Remillard}{McClintock \&
  Remillard}{2006}]{McClintock2006}
McClintock J.~E.,  Remillard R.~A.,  2006, In: Compact stellar X-ray sources.
  Edited by Walter Lewin \&amp; Michiel van der Klis. Cambridge Astrophysics
  Series

\bibitem[\protect\citeauthoryear{Menou, Esin, Narayan, Garcia, Lasota \&
  McClintock}{Menou et~al.}{1999}]{Menou1999}
Menou K.,  Esin A. A.~A.,  Narayan R.,  Garcia M. R. M.~R.,  Lasota J. J.-P.,
   McClintock J.~E. J.~E.,  1999, The Astrophysical Journal, 520, 276

\bibitem[\protect\citeauthoryear{Migliari, Tomsick, Maccarone, Gallo, Fender,
  Nelemans \& Russell}{Migliari et~al.}{2006}]{Migliari2006}
Migliari S.,  Tomsick J.~A.,  Maccarone T.~J.,  Gallo E.,  Fender R.~P.,
  Nelemans G.,    Russell D.~M.,  2006, The Astrophysical Journal, 643, L41

\bibitem[\protect\citeauthoryear{Narayan, Garcia \& McClintock}{Narayan
  et~al.}{1997}]{Narayan1997}
Narayan R.,  Garcia M.~R.,    McClintock J.~E.,  1997, \apjl, 478, L79+

\bibitem[\protect\citeauthoryear{Narayan \& McClintock}{Narayan \&
  McClintock}{2008}]{Narayan2008}
Narayan R.,  McClintock J.~E.,  2008, \nar, 51, 733

\bibitem[\protect\citeauthoryear{Plotkin, Gallo \& Jonker}{Plotkin
  et~al.}{2013}]{Plotkin2013}
Plotkin R.~M.,  Gallo E.,    Jonker P.~G.,  2013, The Astrophysical Journal,
  773, 59

\bibitem[\protect\citeauthoryear{Rau, Greiner \& Filgas}{Rau
  et~al.}{2011}]{Rau2011}
Rau A.,  Greiner J.,    Filgas R.,  2011, \atel, 3140, 1

\bibitem[\protect\citeauthoryear{Rea, Jonker, Nelemans, Pons, Kasliwal,
  Kulkarni \& Wijnands}{Rea et~al.}{2011}]{Rea2011}
Rea N.,  Jonker P.~G.,  Nelemans G.,  Pons J.~A.,  Kasliwal M.~M.,  Kulkarni
  S.~R.,    Wijnands R.,  2011, The Astrophysical Journal, 729, L21

\bibitem[\protect\citeauthoryear{Reynolds \& Miller}{Reynolds \&
  Miller}{2011}]{Reynolds2011}
Reynolds M. T. M.~T.,  Miller J. M. J.~M.,  2011, \apjl, 734, L17

\bibitem[\protect\citeauthoryear{Russell, Fender, Hynes, Brocksopp, Homan,
  Jonker \& Buxton}{Russell et~al.}{2006}]{Russell2006}
Russell D.~M.,  Fender R.~P.,  Hynes R.~I.,  Brocksopp C.,  Homan J.,  Jonker
  P.~G.,    Buxton M.~M.,  2006, Monthly Notices of the Royal Astronomical
  Society, 371, 1334

\bibitem[\protect\citeauthoryear{Shahbaz, Russell, Zurita, Casares,
  Corral-Santana, Dhillon \& Marsh}{Shahbaz et~al.}{2013}]{Shahbaz2013}
Shahbaz T.,  Russell D.~M.,  Zurita C.,  Casares J.,  Corral-Santana J.~M.,
  Dhillon V.~S.,    Marsh T.~R.,  2013, arXiv:1307.0659, p.~12

\bibitem[\protect\citeauthoryear{Steele, Smith, Rees, Baker, Bates, Bode,
  Bowman, Carter, Etherton, Ford, Fraser, Gomboc, Lett, Mansfield, Marchant,
  Medrano-Cerda, Mottram, Raback, Scott, Tomlinson \& Zamanov}{Steele
  et~al.}{2004}]{Steele2004}
Steele I.~A.,  Smith R.~J.,  Rees P.~C.,  Baker I.~P.,  Bates S.~D.,  Bode
  M.~F.,  Bowman M.~K.,  Carter D.,  Etherton J.,  Ford M.~J.,  Fraser S.~N.,
  Gomboc A.,  Lett R. D.~J.,  Mansfield A.~G.,  Marchant J.~M.,  Medrano-Cerda
  G.~A.,  Mottram C.~J.,  Raback D.,  Scott A.~B.,  Tomlinson M.~D.,    Zamanov
  R.,  2004, in {Oschmann, Jr.} J.~M.,  ed., Astronomical Telescopes and
  Instrumentation {<title>The Liverpool Telescope: performance and first
  results</title>}.
International Society for Optics and Photonics, pp 679--692

\bibitem[\protect\citeauthoryear{Str\"{u}der, Briel, Dennerl, Hartmann,
  Kendziorra, Meidinger, Pfeffermann, Reppin \& al.}{Str\"{u}der
  et~al.}{2001}]{Struder2001}
Str\"{u}der L.,  Briel U.,  Dennerl K.,  Hartmann R.,  Kendziorra E.,
  Meidinger N.,  Pfeffermann E.,  Reppin C.,    al. E.,  2001, \aap, 365, L18

\bibitem[\protect\citeauthoryear{Turner, Abbey, Arnaud, Balasini, Barbera,
  Belsole, Bennie, Bernard \& al.}{Turner et~al.}{2001}]{Turner2001}
Turner M.~J.~L.,  Abbey A.,  Arnaud M.,  Balasini M.,  Barbera M.,  Belsole E.,
   Bennie P.~J.,  Bernard J.~P.,    al. E.,  2001, \aap, 365, L27

\bibitem[\protect\citeauthoryear{Wijnands, Homan, Miller \& Lewin}{Wijnands
  et~al.}{2004}]{Wijnands2004}
Wijnands R.,  Homan J.,  Miller J.~M.,    Lewin W.~H.~G.,  2004, \apjl, 606,
  L61

\bibitem[\protect\citeauthoryear{Wijnands, {in 't Zand}, Rupen, Maccarone,
  Homan, Cornelisse, Fender, Grindlay, van~der Klis, Kuulkers, Markwardt,
  Miller-Jones \& Wang}{Wijnands et~al.}{2006}]{Wijnands2006}
Wijnands R.,  {in 't Zand} J. J.~M.,  Rupen M.,  Maccarone T.,  Homan J.,
  Cornelisse R.,  Fender R.,  Grindlay J.,  van~der Klis M.,  Kuulkers E.,
  Markwardt C.~B.,  Miller-Jones J. C.~A.,    Wang Q.~D.,  2006, \aap, 449,
  1117

\bibitem[\protect\citeauthoryear{Wijnands, Yang \& Altamirano}{Wijnands
  et~al.}{2012}]{Wijnands2012}
Wijnands R.,  Yang Y.~J.,    Altamirano D.,  2012, \mnras, 422, L91

\bibitem[\protect\citeauthoryear{Yang, Kong, Russell, Lewis \& Wijnands}{Yang
  et~al.}{2013}]{Yang2013}
Yang Y.~J.,  Kong A. K.~H.,  Russell D.~M.,  Lewis F.,    Wijnands R.,  2013,
  Monthly Notices of the Royal Astronomical Society, 427, 2876

\end{thebibliography}
%\bibliography{/Users/montserrat/Documents/Mendeley/BibTex/library_bib}

\label{lastpage}
\end{document}